\documentclass[12pt,preprint]{aastex}

\newcommand{\lsim}{\raise0.3ex\hbox{$<$}\kern-0.75em{\lower0.65ex\hbox{$\sim$}}}
\newcommand{\gsim}{\raise0.3ex\hbox{$>$}\kern-0.75em{\lower0.65ex\hbox{$\sim$}}}

\shorttitle{Selection of high-$z$ supernovae candidates}
\shortauthors{Dahl\'{e}n \& Goobar}

\begin{document}


\title{Selection of high-$z$ supernovae candidates}


\author{Tomas Dahl\'{e}n}
\affil{Stockholm Observatory, SCFAB, SE-106 91 Stockholm, Sweden
 }
\email{tomas@astro.su.se}
\and
\author{Ariel Goobar}
\affil{Physics Department, Stockholm University, SCFAB, SE-106 91 Stockholm, Sweden}
\email{ariel@physto.se}


\begin{abstract}
Deep, ground based, optical wide-field supernova searches are capable of detecting a
large number of supernovae over a broad redshift range up to
$z~\sim~1.5$. While it is practically unfeasible to obtain
spectroscopic redshifts of all the supernova candidates right
after the discovery, we show that
the magnitudes and colors of the host galaxies, as well as the
supernovae, can be used to select high-$z$ supernova candidates, for
subsequent spectroscopic and photometric follow-up.
 
Using Monte-Carlo simulations we construct criteria for 
selecting galaxies in well-defined redshift bands.
For example, with a
selection criteria using $B-R$ and $R-I$ colors we are able
to pick out potential host galaxies for which $z~\ge~0.85$ with 
80\% confidence level and with a selection efficiency of
64--86\%. The method was successfully tested using 
real observations from the HDF. 

Similarly, we show that that the
magnitude and colors of the supernova discovery data can be used to constrain the
redshift. With a set of cuts based on $V-R$ and $R-I$ in a search to
$m_I~\sim~25$, supernovae at $z \sim 1$ can be selected in 
a redshift interval  $\sigma_z \le 0.15$.

\end{abstract}




\section{Introduction}
High-$z$ Type Ia supernovae (hereafter SNe) have been shown to be very
accurate tools for studying cosmological parameters
\citep{per99,rie98}. As the sensitivity of the magnitude-redshift
relation increases with the size of the redshift range probed,
searches for SNe at redshifts $z~\gsim$~1 become crucial to refine
our understanding of cosmology \citep{goo95,gol01}.

One of the difficulties in searches dedicated to high-$z$ SNe is to
select candidates for subsequent follow-up. An $I$-band SN search down
to $m_I~\sim~25$ generates candidates in a broad redshift range:
$0.1~<~z~\lsim~1.5$. On a 1 square degree detection image,
separated by three weeks from a reference image, there will be about a
 dozen Type Ia SNe on the rising part of the light curve.

Additional information is required to make the follow-up of
$z~>$ 0.85 SNe efficient. This is particularly important as the
follow-up of candidates with space instruments, such as
HST, needs to be decided upon very soon after discovery, leaving
limited time for prior spectroscopic screening.
  
In this paper we discuss how broad-band photometry of the host galaxy,
as well as the SN can be used in order to select targets with
$\sigma_z \sim 0.15$ with high efficiency. We focus on
the optical bands, producing mainly SNe up to $z~\sim~1.5$. In Section 2 
we discuss the SN rates that are used in our
simulations. In Section 3 we study the use of host galaxy magnitudes
and colors to estimate the redshift of a SN candidate, followed by a
discussion in Section 4 on how deep a survey must be in order to
obtain the required galaxy colors. In section 5 we discuss the use of
SN magnitude and color for redshift estimation. Finally, a summary and
conclusions are given in Section 6.

In this paper we assume a flat cosmology with $\Omega_M$ = 0.3 and
$\Omega_{\Lambda}$ = 0.7, and a Hubble constant $H_0 = 65$ km s$^{-1}$
Mpc$^{-1}$ Magnitudes are given in the standard Vega based system.
\section{SN rates}
Models of high-$z$ SN rates predict an increase out to $z~\sim$ 1.5
\citep{mad98a,dah99}. This increase reflects the observed increase in
the star formation in galaxies out to these redshifts. (e.g. Madau,
Pozzetti \& Dickinson 1998). At higher redshifts the evolution of the
star formation rate, and therefore SN rate, is more uncertain. This is
mainly due to the uncertainty in the amount of dust extinction
affecting observations of high-$z$ galaxies. A direct relation between
star formation rate and SN rate is, however, only valid for core
collapse SN (Types II, Ib/c), where the SN progenitor is a short lived
star. For Type Ia SNe there is an unknown time delay between
the formation of the SN progenitor star and the explosion of the
SN. In the models cited above, a parameter $\tau$ is introduced to
represent this time delay. Considered values of $\tau$ lie in the
range 0.3-3 Gyr.
  
Calculating the absolute numbers for the SN rates involves a number
parameters, such as the distribution of SN types, peak luminosities,
light curves, IMF, cosmology, and most important, assumptions about
dust extinction. Estimated rates could therefore have an uncertainty
of about a factor 2-3. Using simulations described in \cite{dah99}, we
estimate that a detection image, taken three weeks after a reference
image, should result in $\sim~20-45$ SNe per sq. deg. Here we have used
a limiting magnitude $m_I~=~25$, and a detection criteria where the SNe
on the second image are $>~0.5$ mag brighter than on the reference
image. Approximately half of the SNe are Type Ia that are still brightening.

In this paper we are not directly affected by the absolute value of
the SN rate at different $z$, since the aim here is to estimate the
redshift for SN candidates already detected. However, in our
simulations we estimate the redshift {\em distribution} of the
observable SNe, therefore, the relative evolution of the SN rate with
$z$ must be considered. In our models we use both a constant and an
evolving Type Ia SN rate, where the latter is taken from the
theoretical models in \cite{dah99}. 

\begin{figure}
\figurenum{1}
\epsscale{0.8}
\plotone{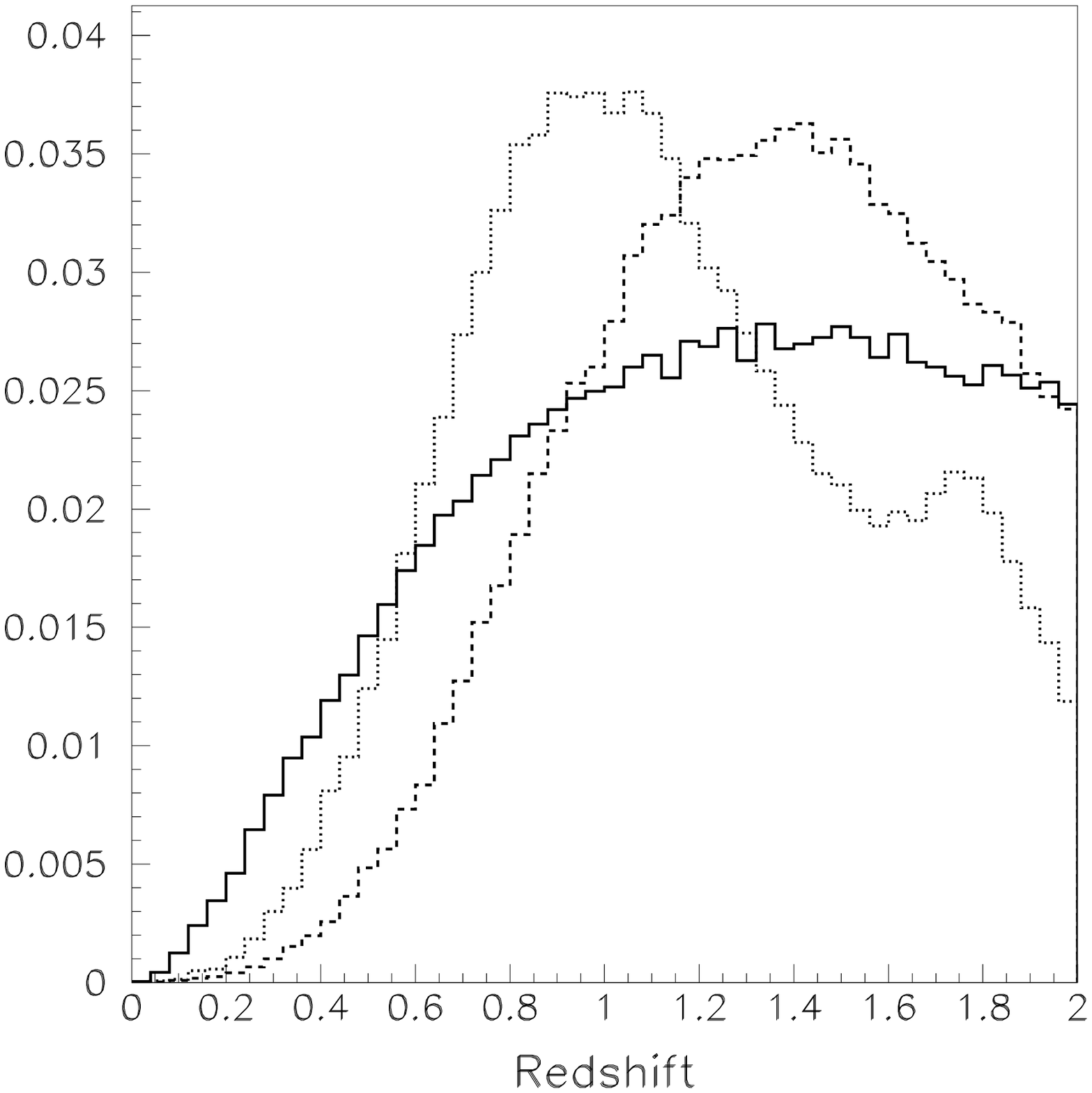}
\caption{Differential observable Type Ia SN rate as a function of redshift for
three models (arbitrary units), were the integrated rate up to $z~=~2$ has been normalized
to 1 using 10$^5$ MC simulated SNe.
The solid line assumes a constant SN rate per comoving
volume. The dashed line shows the observable rate for $\tau~=~1$ Gyr,
the dotted line $\tau~=~3$ Gyr. Note that apart from the volume effect,
the observable rate falls as $(1+z)^{-1}$ due to the cosmological time
dilation.}
\label{fig1}
\end{figure}

Figure \ref{fig1} shows the differential observable Type Ia SN rate vs redshift 
(arbitrary units) for models with constant SN rate (solid line) and evolving rates with
$\tau~=~1.0$ Gyr (dashed line) and $\tau~=~3.0$ Gyr (dotted line). The
distribution was generated with a Monte-Carlo (MC) simulation of $10^5$
SNe (for each model) up to a redshift $z~=~2$. The integrated rate is
normalized to unity for all three cases.

Note that the shape of the distribution is given by two effects
when going from intrinsic rates, to rates in the observer's frame: the
volume effect makes the number of high-$z$ SNe increase to
$z~\sim~1.8$, and the cosmological time dilation makes the rates fall
as $(1+z)^{-1}$.
\section{Selection by host galaxy colors}
Using the magnitude and colors of the host galaxy to select SN
candidates has the advantage that a catalog listing probable high-$z$
galaxies can be constructed in advance. When a SN is detected, it
possible to immediately determine if it is a high-$z$ candidate. Also,
the host galaxy redshift gives an estimate of the absolute magnitude
of the SNe, which can set constraints regarding SN type, since Type
Ia's have brighter peak luminosities than other SN types, except for a small fraction of extremely luminous core collapse SNe. This is
important since an increasing fraction of the SNe will be Type Ib/c or
II as one reaches for fainter magnitude limits. 

Using host galaxy magnitudes requires deep images of the survey field from which colors can be obtained. The limiting magnitude of this field should typically be $\gsim~1$ mag deeper than the limiting magnitude of the SN search images in order to reach equivalent depth for the extended galaxies and the point source SNe. A natural way of obtaining a deep background field is to add the images from the subsequent search epochs.

When calculating the galaxy properties we are here primarily interested in the host galaxies of Type Ia SNe. Therefore, we include all types of galaxies, even though only Type Ia's have been found to occur in ellipticals (e.g. Wang, H\"{o}flich \& Wheeler 1997).

The accuracy of redshift determination by using host galaxy
magnitude and colors depend on the limiting magnitude, the accuracy in
the photometry and the choice of filters, and is a trade off between the required precision in redshift and the amount of observing time available.
\subsection{One filter}
Using the host galaxy magnitude in the band where the SN search is
conducted can be used to calculate a rough probability of the SN being
at high $z$. To quantify this we estimate the redshift distribution of
galaxies as a function of the observed $I$-band magnitude. We simulate
the field by distributing galaxies over a redshift range $0~<~z~<~2$,
with absolute magnitudes according to a field luminosity function
(LF). We choose a LF described by two Schechter functions
\citep{sch76}, one for early type galaxies with $M_B^*=-20.2$ and
$\alpha=-0.8$, and one for late type galaxies with $M_B^*=-20.6$
and $\alpha=-1.25$. These values are consistent with results from
\cite{met91}. Note, however, that the uncertainties in these
parameters, especially $M_B^*$, are fairly large, and it is not clear
if early-types or late-types have the brightest $M_B^*$ (e.g. Metcalfe
et al. 1991; Lilly et al. 1995; He et al. 2001). It is also expected
that the LF evolve with redshift \citep{lil95}. At first we do not
include a luminosity evolution, but take this into account further
down. 

The two Schechter functions are normalized so that the fraction of early-type galaxies is 15\%. The late-type galaxies are further divided so that 35\% of the total number of galaxies are spirals and 50\% Irregulars. This is
approximately the same fractions as for galaxies in the HDF to
$z~\sim$ 1 (Fern\'{a}ndes-Soto, Lanzetta \& Yahil 1999).

\begin{figure}
\figurenum{2}
\epsscale{1.0}
\plotone{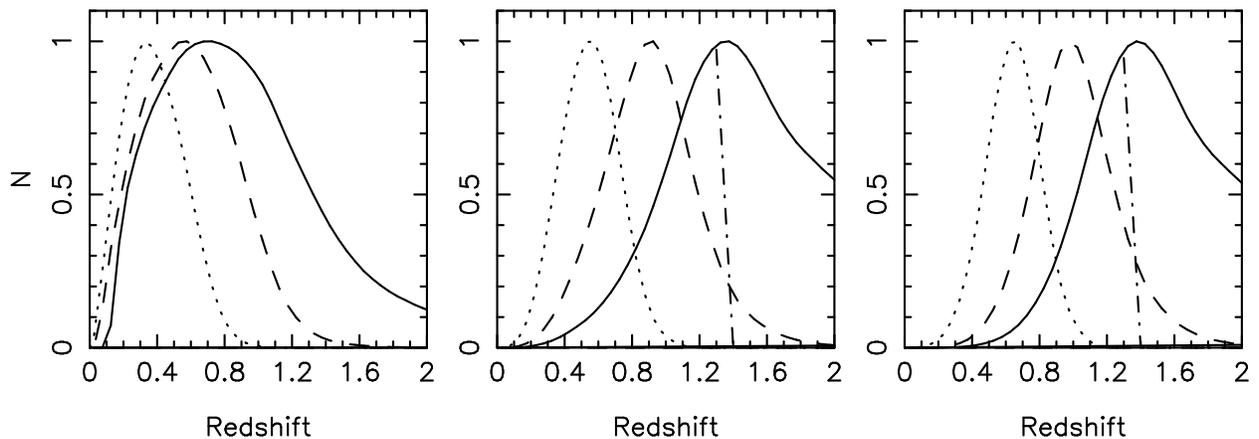}
\caption{The left panel shows the redshift distribution of field galaxies in three different magnitude intervals, $20 < m_I < 21$ (dotted line), $22 < m_I < 23$ (dashed line) and $24 < m_I < 25$ (solid line). The middle panel shows the redshift distribution of SN host galaxies. We have here weighted each galaxy with its intrinsic $B$-band luminosity in order to take into account that bright galaxies have higher probability of hosting a SNe. The right panel shows the distribution of SN host galaxies where we assume an evolving SN rate with redshift. The dash-dotted line represents a high-$z$ cut-off at $z~\sim~1.4$, since in a $m_I~\lsim~25$ search, SNe are too faint to be detected at higher redshifts. Therefore, there should be no host galaxies beyond this cut-off. The peak in each distribution is normalized to unity.}
\label{fig2}
\end{figure}

Observed magnitudes are calculated by adding the distance modulus and
type dependent K-corrections to the rest frame
magnitudes. K-corrections are calculated from the E, Sbc, Scd and Im
galaxy templates in \cite{col80}. We also include a Sa template from
\cite{kin96}. In the left panel of Figure \ref{fig2} we plot the
resulting redshift distribution for three magnitude bins, $m_I$ =
20--21 (dotted line), $m_I$ = 22--23 (dashed line) and $m_I$ = 24--25
(solid line). The mean redshifts in these bins are $\bar{z}$ = 0.38,
0.59 and 0.75, respectively. The percentages of galaxies with redshift
$z~>$ 0.85 are 0.6, 18 and 39\%, respectively. Without any additional
information these results indicate that a host galaxy should have
$m_I~>~24$, in order to have a reasonably high probability of being at
high redshift. Note that the peak in each distribution in Figure \ref{fig2}
is normalized to unity.

Next we take into account that SNe should be more frequent in galaxies
with high intrinsic luminosity. We do this by weighting each galaxy
with its rest-frame $B$-band luminosity. For Type Ia's, with the unknown time delay
between star formation and SN explosion, the relation between galaxy luminosity and SN rate is probably more complex than a pure proportionality. However observations show that Type Ia SN rates, as well as Type II rates, are close to proportional to the $B$-band luminosity \citep{cap93}. Therefore, in order to compensate for galaxy luminosities, we use the assumption that Type Ia rates are proportional to $B$-band luminosities.

\begin{figure}
\figurenum{3}
\epsscale{1.0}
\plotone{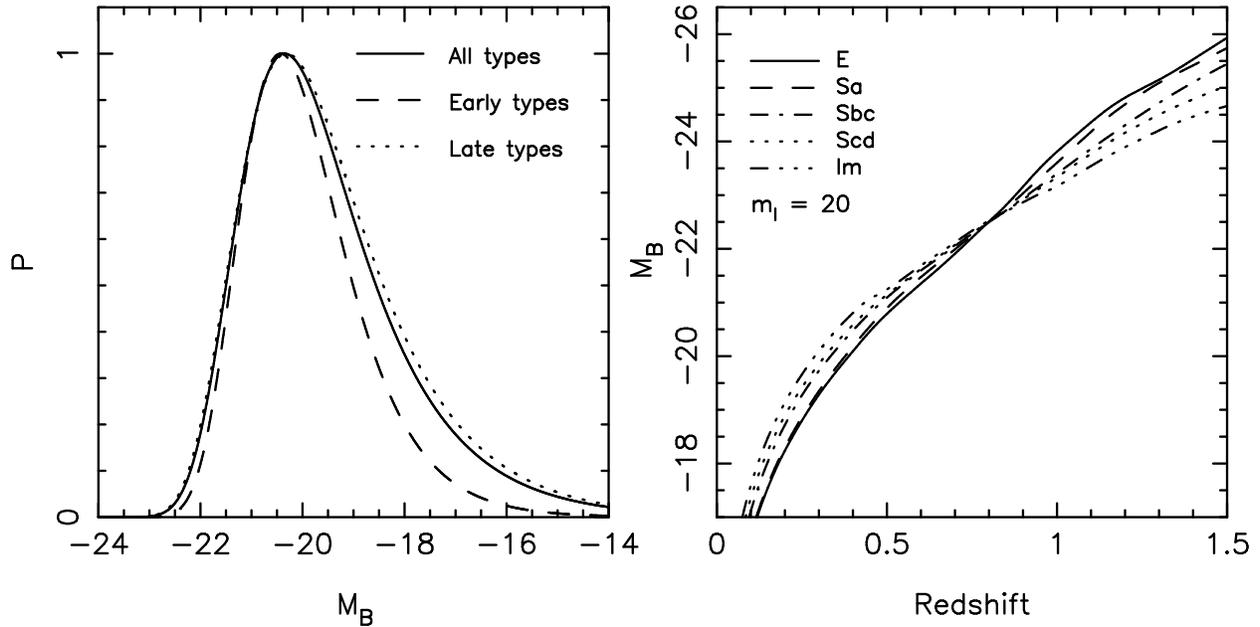}
\caption{The left panel shows the probability distribution of SN host galaxy absolute magnitudes in the $B$-band. The solid line shows the average probability for all galaxy types, while the dashed and dotted lines show the probability for early-types and late-types separately. The distributions are derived by weighting the $B$-band luminosity function with the intrinsic luminosity at each magnitude. Peaks are normalized to unity. The right panel shows $M_B$ as a function of redshift for a galaxy with observed apparent magnitude $m_I$ = 20, for five different galaxy types.}
\label{fig3}
\end{figure}

In the left panel of Figure \ref{fig3} we show the resulting
probability distribution for host galaxy absolute magnitudes. The
solid line shows the total distribution, while the dashed and dotted
lines show the distribution divided into early-type and late-type
galaxies, respectively. The most probable host galaxy magnitudes lies
in a range of $\Delta M~\pm\sim$ 1.5 mag from $M_B^*$. Note also that
the shape of the LF creates a sharp cut-off at bright magnitudes. This can be
used in a first rejection of host galaxies based on their apparent
magnitudes. For example, a galaxy with apparent magnitude
$m_I~\sim~20$ would have to have $M_B~\lsim~-22.5$, in order to be at
$z~\gsim~0.8$, which according to the plot is unlikely. A strongly
evolving LF could, however, move this cut-off to brighter
magnitudes. In the right panel of Figure \ref{fig3} we plot $M_B$ as a function of redshift for a galaxy with observed magnitude $m_I~=~20$, for five different galaxy types. The choice of magnitude, $m_I~=~20$, is here arbitrary, the plot can be scaled to any apparent $I$-band magnitude.
 
Using the probability function for host galaxy magnitudes, we plot the
redshift distribution of SN host galaxies in the middle panel of
Figure \ref{fig2}, using the same magnitude intervals as above. The
distributions are here shifted towards higher redshift compared to the
left panel. This is natural since intrinsically bright galaxies can be
detected to higher redshifts. We must also consider that SNe are not
likely to be detected at $z~\gsim~1.4$ in optical searches to
$m_I~\sim~25$ (shown in a later section). Therefore, we may safely neglect
host galaxies at these redshifts. To compensate for this we
introduce a linear cut-off between $z~=~1.3$ and $z~=~1.4$, as
illustrated this by the dash-dotted line in Figure \ref{fig2}. The
mean redshifts in the three magnitude intervals, $m_I=$ 20--21, 22--23
and 24--25, are $\bar{z}$ = 0.54, 0.87 and 1.06, respectively. The
percentages of probable host galaxies with $z~>~0.85$ are 3, 54 and
82\% for the three magnitude bins, indicating that $m_I~\sim~22$
should be sufficiently deep for a relatively high probability of a
host galaxy being at high $z$.

So far we have used a non evolving LF, which here is equivalent to assuming
a constant SN rate with redshift. Next we take the increase in galaxy
luminosity at high-$z$ into account by using an evolving Type Ia SN
rate ($\tau~=~1$ Gyr model from \cite{dah99}). The resulting host
galaxy distribution is shown in the right panel of Figure
\ref{fig2}. There is here a small shift towards higher redshifts
compared to the case with constant SN rate. The mean redshifts in the
three magnitude intervals are now $\bar{z}$ = 0.64, 0.96 and 1.12. The
fractions of host galaxies with $z~>~0.85$ in the three bins are 9, 70
and 90\%, respectively.

\begin{figure}
\figurenum{4}
\epsscale{1.0}
\plotone{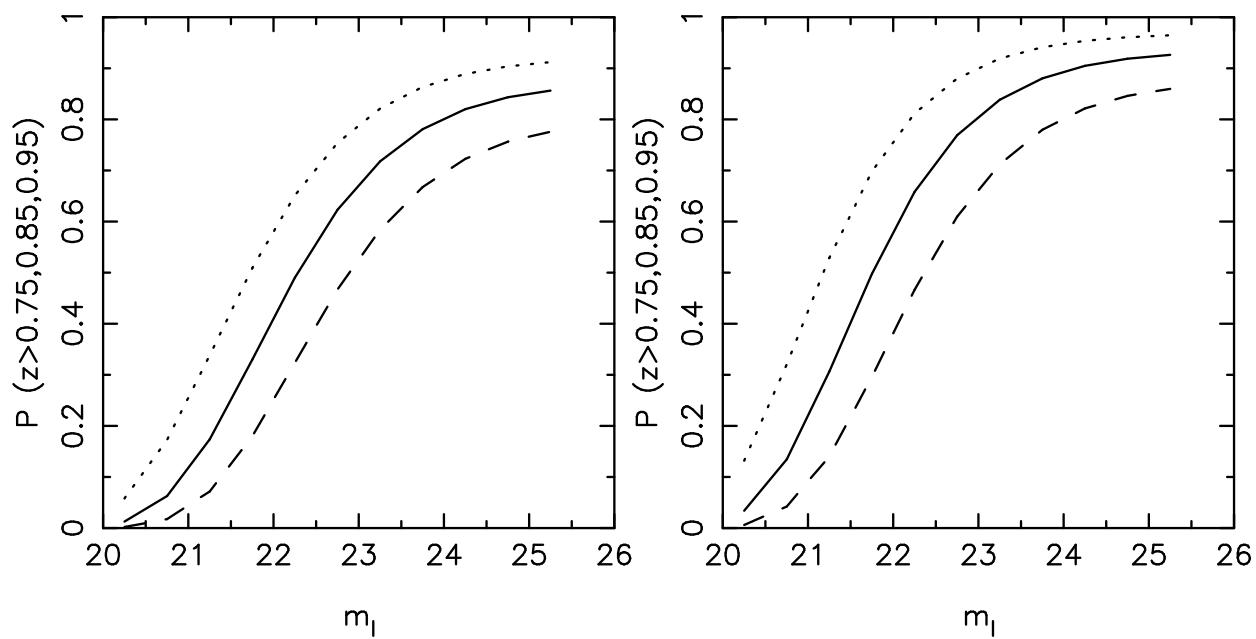}
\caption{The probability that a SN host galaxy has $z~>$ 0.75 (dotted line), $z~>$ 0.85 (solid line) and $z~>$ 0.95 (dashed line) as a function of observed $I$ magnitude. The left panel shows the case for a constant SN rate, while the right panel shows the case for a SN rate that evolves with redshift.}
\label{fig4}
\end{figure}

In Figure \ref{fig4} we plot the probability that a host galaxy has a
redshift $z~>~0.75$ (dotted line), $z~>~0.85$ (solid line) and
$z~>~0.95$ (dashed line), as a function of apparent $I$-band
magnitude. The host galaxies are here weighted with their intrinsic
luminosities as above. We have used a constant SN rate in the left
panel, and an evolving rate in the right panel.

In the following discussion of host galaxy colors it is important to
take into account these results regarding the apparent magnitudes when
estimating the redshift of SN host galaxies.
\subsection{Two filters}
\begin{figure}
\figurenum{5}
\epsscale{1.0}
\plotone{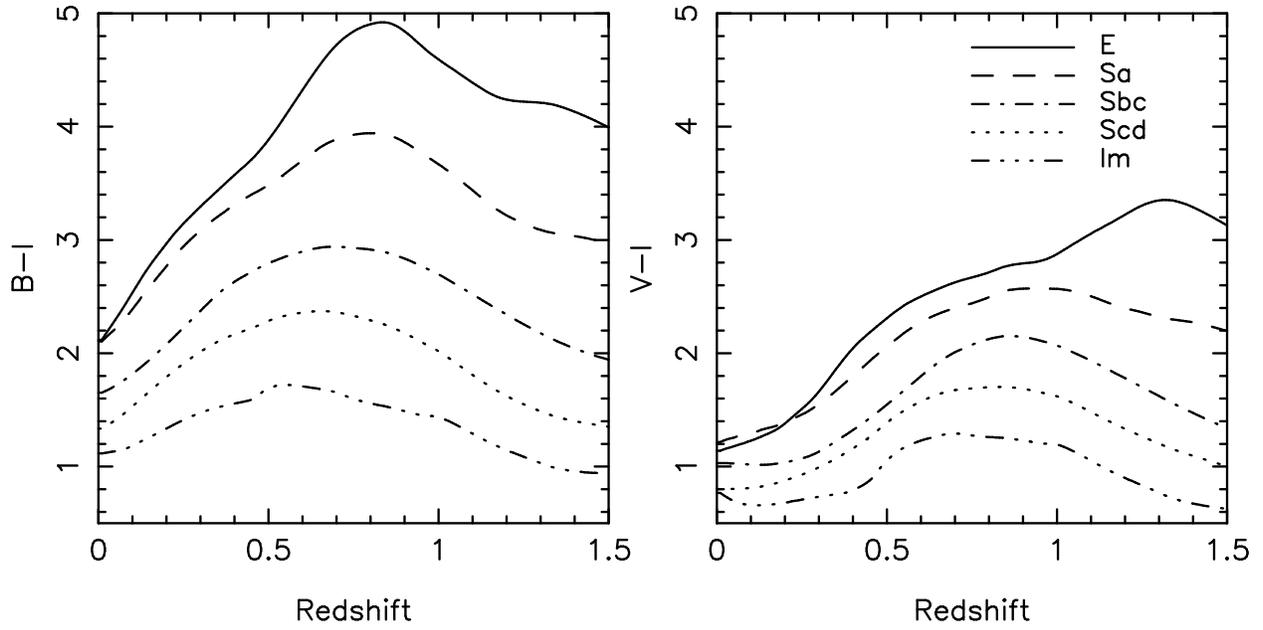}
\caption{$B-I$ and $V-I$ color as a function of redshift for five different types of galaxies.}
\label{fig5}
\end{figure}

With two filters we can use the color information of the host galaxy
together with the apparent magnitude to better constrain the
redshift. In addition to the $I$-band, we here assume that $B$ or
$V$-band magnitudes are available. Figure \ref{fig5} shows $B-I$ and
$V-I$ colors as a function of redshift for the five different galaxy
types. The figure clearly illustrates the strong evolution of the
galaxy color with redshift. The main reason for the reddening with $z$
is the cosmological shift of the 4000\AA ~break out of the $B$- and
$V$-bands at increasing redshift. At $z~\sim~1$ the 4000 \AA \ break
moves into the $I$-band, and the curves flatten. The stronger break in
early-type galaxies leads to a more pronounced color evolution in E
and Sa galaxies. A problem when using a single color is the large
spread between different galaxy types. Morphological information on
the host galaxy makes the selection efficiency better, even though one should remember that there is no one-to-one correspondence between the morphological type and the spectroscopical type giving the colors (e.g., Dressler et al. 1999; Poggianti et al. 1999). Also, the presence of internal redding can affect the colors of a specified morphological type. Therefore, one should be careful when using a single color to estimate redshift.

Note also that the evolution is generally stronger in
$B-I$. However, this also leads to a rapid fainting of the galaxies in
$B$, and long exposure times are needed in order to measure the colors.
\subsection{Three filters}
\begin{figure}
\figurenum{6}
\epsscale{1.0}
\plotone{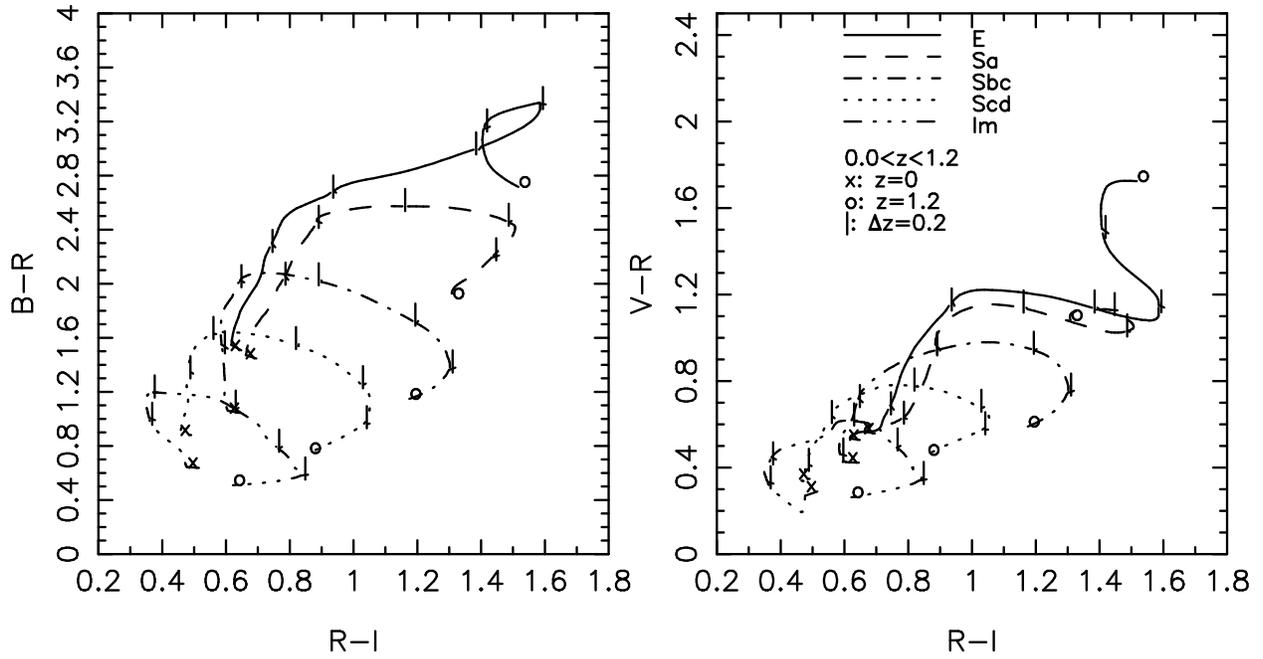}
\caption{Left panel: the evolution of the $V-R$ vs $R-I$ color-color diagram with redshift for five different galaxy types. Right panel: the same for $B-R$ vs $R-I$. High-$z$ galaxies occupy the bottom right part of the plot.}
\label{fig6}
\end{figure}

When the galaxy spectrum is cosmologically redshifted the colors
change, as described in the previous section. This effect can be
quantified by the evolution of observed colors in a color-color
diagram. The left panel of Figure \ref{fig6} shows the position of
galaxies (E, Sa, Sbc, Scd and Im) in a two-dimensional color-color
diagram of $B-R$ vs. $R-I$, while the right panel shows the same for
$V-R$ vs. $R-I$.  The 'x's define the $z~=~0$ position for each galaxy
type. The contours show the color-color evolution up to $z~=~1.2$,
marked by an 'o', where steps in $z$ of 0.2 are shown by a '$|$' mark.

Note that marks defining the same redshifts approximately fall on straight "iso-redshift" contours going from the bottom left to the top right part of the figures. Including a larger sample of templates with intermediate types results in color-color curves with redshift marks approximately falling on these contours. Including internal redding in the templates also shifts the curves along the "iso-redshift" contours. E.g., assuming $E(B-V)~=~0.3$ in an Scd galaxy makes the colors become almost identical to an Sbc galaxy. Internal redding is therefore not a severe problem when estimating redshifts using color-color selection.

As can be seen in the plot, the bottom right part defines a region
with $z~\gsim~0.85$. Note also that using $B$ instead of $V$ results
in a more distinct evolution (note different scales on the y-axis). As
already mentioned, however, the trade-off with the $B$-band is that
significantly more observing time is required. We discuss this in
Section 4.

One should notice that there are many factors involved when using
color-color selection, i.e filter choices, magnitude limits, accuracy
in magnitudes, the galaxy types of interest and their absolute
magnitudes. To quantify how three filter color-color selection can be
used in order to select high-$z$ candidates we use MC simulations. In
the example we choose to select host galaxies with $z~\ge~0.85$. For
this redshift limit, $R$ and $I$ filters should be used, since here
the 4000 \AA \ break has just left the $R$-band and is moving into the
$I$-band. In the simulations, both the $B$- and $V$-bands are used as
the third filter. The $U$-band has not been considered due to the long
exposure times that would be needed in that case.
\subsubsection{Color--color selection with $BRI$ and $VRI$}
A total of $\sim$ 250000 galaxies were generated and distributed from
early-type ellipticals to irregulars, with a distribution
according to the LF described in Section 3.1, and in a redshift range
$0~<~z~<~1.4$. Observed magnitudes are calculated by adding distance
modulus and type dependent K-corrections. To each magnitude (i.e $B$,
$V$, $R$ and $I$) of each galaxy, a random error is generated as to
mimic the real uncertainties in photometry. We use a normal
distribution with a sigma value varying from 0.06 mag at bright
magnitudes, $m_I~<$ 20, to 0.1 mag at our limit $m_I$ = 25. 
Besides the variation in colors due to spectral type and internal redding, there is also a possibility of an intrinsic variation in the colors within a specified spectral type. To account for this we add an extra dispersion of 0.1 mag to each color. This is consistent with the difference in colors between the template sets of \cite{col80} and \cite{kin96}, when comparing elliptical and Sbc galaxies individually. We distribute galaxies out to the cut-off at $z~\sim~1.4$, described in Section 3.1. We consider two models with different SN rates. The first model has a constant SN rate, equivalent with a non-evolving LF. In the second model we weight each galaxy with the SN rate taken from the evolving model with $\tau~=~1$.

\begin{figure}
\figurenum{7}
\epsscale{1.0}
\plotone{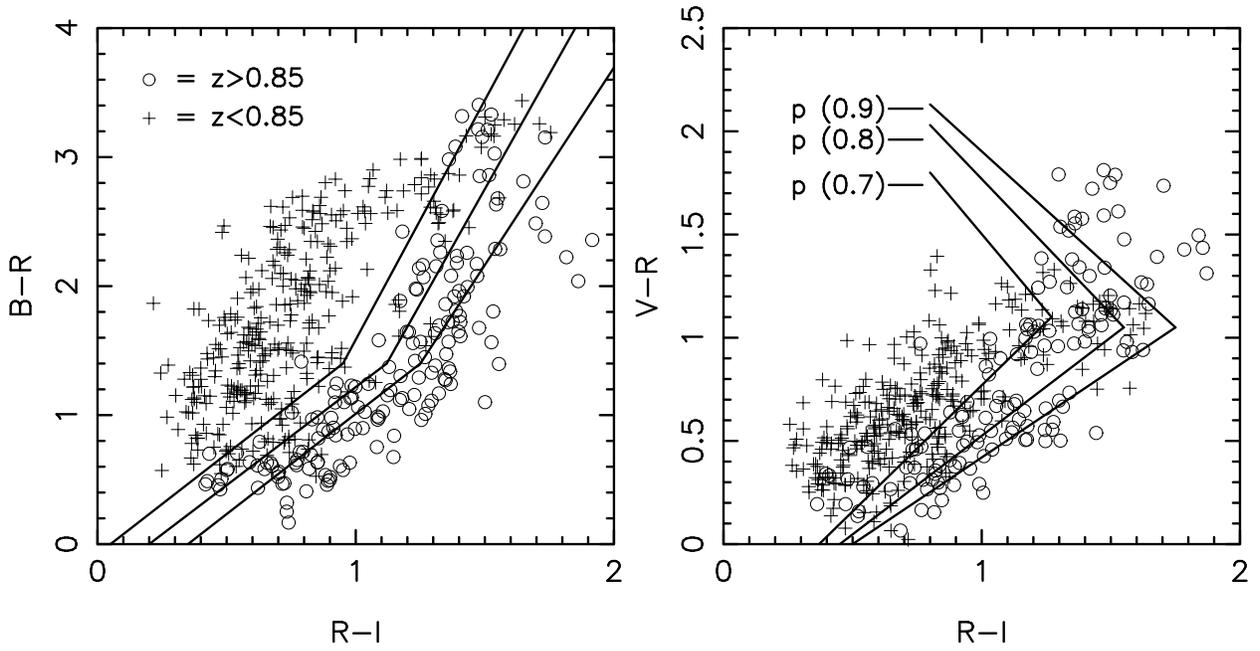}
\caption{Left panel: Color-color selection criteria in $B-R$ vs $R-I$. Crosses represent galaxies with $z~<~0.85$, and circles galaxies with $z~>~0.85$. On the right side of the lines the probability that a host galaxy has $z~>$ 0.85 is 0.7, 0.8 and 0.9 for the left, middle and right line, respectively. Right panel: the same for $V-R$ vs. $R-I$.}
\label{fig7}
\end{figure}

Figure \ref{fig7} shows a representative fraction of the generated
galaxies in a color-color diagrams in the case of a non-evolving
LF. The circles represent galaxies with redshift $z~>~0.85$, crosses
show the galaxies with $z~<~0.85$. Three lines are also shown in the
figure as examples of selection criteria. On the right side of these
lines the probability that a host galaxy has $z~>$ 0.85 is 0.7, 0.8
and 0.9 for the left, middle and right lines, respectively. In the
following we call these probability criteria $p_{(0.7)}$, $p_{(0.8)}$
and $p_{(0.9)}$. As expected, these criteria are almost parallel to the "iso-redshift" contours for $z~\sim~0.85$. Therefore, using a larger set of templates, or including internal redding do not affect the results since this mainly moves the galaxy colors along these contours.

The difference in selection efficiency between using $B$ and $V$ can
be quantified by calculating the fraction of the simulated galaxies
with known redshift $z~>~0.85$, that are actually picked out by the
selection criteria. With $BRI$ selection, (40,64,86)\% of the galaxies
that have $z~>$ 0.85 are picked out when using
($p_{(0.9)},p_{(0.8)},p_{(0.7)}$). With $VRI$ selection, the numbers
are (31,47,73)\%. For the model with evolving SN rate, there are
relatively more SNe at high $z$, therefore the selection efficiency
is higher in this model. We find that (51,86,99)\% of the
galaxies that have $z~>$ 0.85 are picked out when we construct
selection criteria ($p_{(0.9)},p_{(0.8)},p_{(0.7)}$) in $BRI$. With
$VRI$, the percentages are (43,80,98)\%.

Similar selection criteria can easily be produced for other choices of
redshift, filters, photometric errors and limiting magnitudes. Note
that it is important that the selection criteria are calculated for a
limiting magnitude matching that of the real observations. For
example, if the limiting magnitude is $m_I$ = 22, instead of $m_I$ =
25 as used here, then the lines representing the 0.9 probability
criterion in the plot, will no longer be valid. Instead, these lines
would represent a $\sim$ 0.6--0.7 probability.

Note also that the apparent magnitudes for the individual galaxies are
not explicitly used in the color-color selections (except that they
must be brighter than the limiting $I$ magnitude). Therefore, one
should take the results regarding the absolute magnitudes of the host
galaxy into account, e.g. the results described by Figure \ref{fig3}.
\subsubsection{Checking the simulation against HDF}
\begin{figure}
\figurenum{8}
\epsscale{0.7}
\plotone{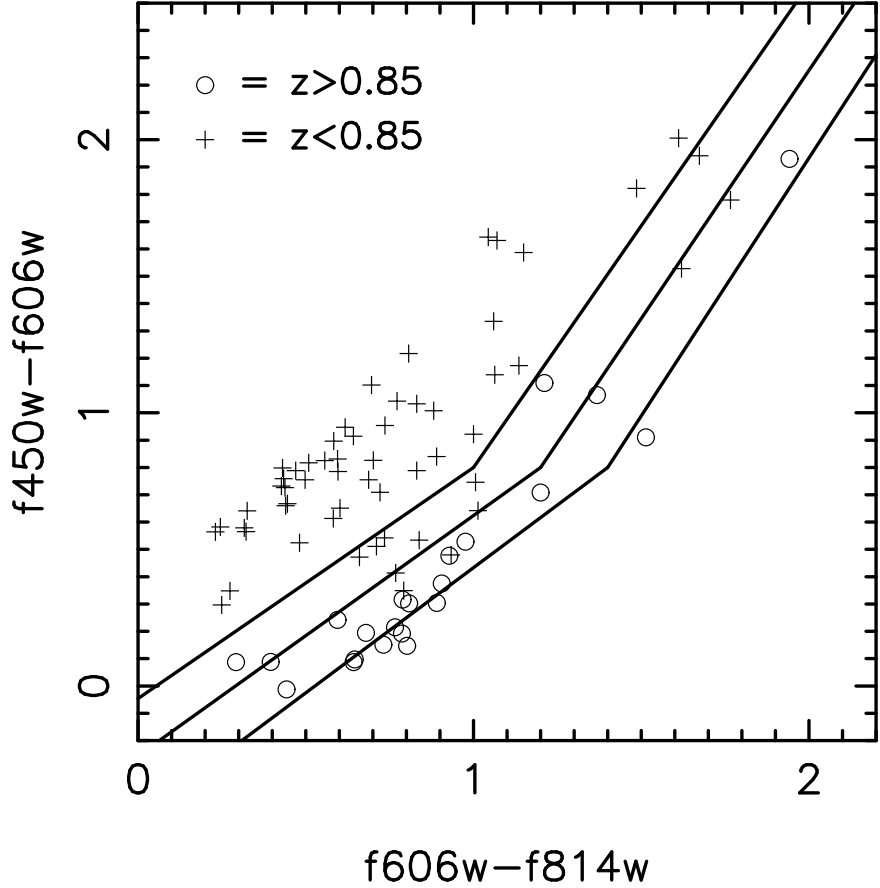}
\caption{Color-color selection criteria for the HDF calculated from MC simulations. To the right of the three lines the probabilities are 0.7, 0.8 and 0.9 that a galaxy has $z~>~0.85$, going from left to right. Galaxies in the HDF with spectroscopic redshift $0.85~<~z~<~1.4$ are marked by circles, while galaxies with $z~<~0.85$ are marked by crosses. Spectroscopic and photometric data for the HDF are taken from \cite{fer99}.}
\label{fig8}
\end{figure}

To test the color-color selection against real observations we
calculate selection criteria for the HDF-N using MC simulations as
above, using a constant SN rate. We use the f450w, f606w and f814w
filters, corresponding to $B$, $V$ and $I$. In
Figure \ref{fig8} we plot selection criteria analogue to Figure
\ref{fig7}, i.e to the right of the three lines the probabilities are
0.7, 0.8 and 0.9, that a host galaxy has $z~>~0.85$, going from left
to right lines. In the figure we have also plotted galaxies in the HDF
with spectroscopic redshift $0.0~<~z~<~1.4$ and $m_I~<~25$. Photometry
and redshifts for these galaxies are taken from the catalog published
by \cite{fer99}. Galaxies with spectroscopic redshift $z~>~0.85$ are
marked by circles, while lower redshift galaxies are marked by
crosses. In total there are 79 galaxies to the limiting magnitude, of
which 22 are at $z~>~0.85$. The figure shows that the high-$z$ and
low-$z$ galaxies occupy different regions in the color-color plot. To
the right of the $p_{(0.7)}$ line, 65\% of the real galaxies have
$z~>~0.85$. For the $p_{(0.8)}$ and $p_{(0.9)}$ lines, 80 and 100\% of
the galaxies have $z~>~0.85$. Of the total number of galaxies with
$z~>~0.85$, 100\% lie to the right of $p_{(0.7)}$, while 91 and 36\%
lies to the right of $p_{(0.8)}$ and $p_{(0.9)}$, respectively.
\subsection{Four or more filters}
With a set of four or more filters available, it is possible to
set {\em both} lower and upper limits on the redshifts of individual galaxies, using photometric
redshift techniques (as compared to the three filters case where we
estimate the probability that a galaxy has a redshift above, {\em or} below,
some chosen limit). The photometric redshift technique has developed
much during the last years, and has now been used in a variety of
applications. There are basically two different approaches to the
technique. In the empirical fitting method (e.g. Connolly et al. 1995;
Connolly et al. 1997; Wang, Bahcall \& Turner 1998; Brunner, Connolly
\& Szalay 1999) a training set of galaxies with known redshifts are
used to derive a polynomial giving the redshift as a function of
colors and magnitudes. In the template fitting technique
\citep{pus82,gwy95,mob96,saw97,fer99,bol00,fon00} a chi-square fit is
made between observed colors and as set of galaxy templates redshifted
over a range of redshift. The accuracy of the method to $z~\lsim~1.3$
can reach $\sigma_z \sim 0.05(1+z)$ \citep{coh00}, depending on the
number of filters and photometric errors.  With this accuracy,
high-$z$ host galaxies can readily be found. The trade-off with the method is,
of course, the large amount of observing time required to map a large sky coverage inat least four bands. A public available code, {\it hyperz}, for
calculating photometric redshifts using the template fitting method is
published by Bolzonella et al. (2000).

\section{Galaxy magnitudes}
In this analysis we have used a limiting magnitude $m_I~=~25$. Two
questions should be addressed here. Is this deep enough to reach host
galaxies at redshifts of interest, and what limiting magnitudes are
required in the other filters in order to get the desired colors? In
Table \ref{Table1} we give the rest frame $B$ magnitudes as a function
of redshift for different galaxy types, assuming an observed magnitude
$m_I~=~25$. Comparing these results with Figure \ref{fig3} shows that
most host galaxies at $z~\lsim~1$ should be reached, since we here
reach an absolute magnitude $M_B~\gsim~-19$. To quantify this further
we also list the fraction of SN host galaxies that can be detected at
different redshifts, assuming three different limiting magnitudes,
$m_I~=~24$, $m_I~=~25$ and $m_I~=~26$.

Next we investigate the limiting magnitudes that must be reached in
different filters to get the colors of galaxies of different
types, and at different redshifts. In Table \ref{Table2} we list the
apparent magnitudes in $B$, $V$ and $R$ for a galaxies with an {\em
observed magnitude $m_I$ = 25}. We list the magnitudes for E, Sbc and
Im galaxies. Sa and Scd galaxies have magnitudes approximately in
between these.

Note that Table \ref{Table2} does not imply that galaxies get brighter
at higher redshift. If we consider two elliptical galaxies at, i.e.,
$z$ = 0.8 and $z$ = 1.4, that have the {\em same apparent magnitude in
the $I$-band}, then the galaxy at $z$ = 1.4 will be 0.8 mag brighter
in $B$, than the galaxy at $z$ = 0.8. This is a consequence of the
turn-over at $z~\gsim~0.8$ in the $B-I$ color vs. redshift diagram
(Figure \ref{fig5}).

Finally we estimate the limiting magnitudes in $B$, $V$ and $R$ that
are required in order to detect 70, 80, 90 and 95\% of the host
galaxies with $m_I~\le~25$, in these three filters, at different
redshifts. Results are listed in Table \ref{Table3}. Again, in some
bands there is an apparent trend of brighter limiting magnitudes at
higher $z$, which is explained by the use of a constant limit on the
$I$-band. One must here take into account that a decreasing fraction
of the total number of host galaxies are detected to this limit in
$I$, as shown in the "$f(m_I~=~25)$" column of Table \ref{Table1}.
 
It is also important to remember that the fractions given in Table
\ref{Table1}, and the magnitudes given in Table \ref{Table3}, are
calculated for SN host galaxies, i.e. where we have weighted each
galaxy with its intrinsic luminosity. The fractions and magnitudes are
therefore not valid for a pure galaxy search.

As an example, Table \ref{Table3} shows that in order to get colors of
90 percent of the objects at $z~\sim~1$, detected in a survey with
limiting magnitude $m_I~=~25$, it is necessary to reach
$m_B~\sim~28.1$, $m_V~\sim~26.8$ and $m_R~\sim~25.6$. This information
should be complemented with the results in Table \ref{Table1}, which
shows that $\sim~82$\% of the total number of host galaxies at
$z\sim~1$ should be brighter than the limiting magnitude
$m_I~=~25$. The typical telescope time needed to reach $m_I~=~25$ with
an 8m class telescope is $\sim~0.3$h (1.1h) for a S/N~=~5 (10). The
total time required to get three color information in $VRI$ is
$\sim~1.5$h (6h), assuming depths $m_V~\sim27$, $m_R~\sim~26$ and
$m_I~\sim~25$. Adding the $B$-band requires an additional 2.2h (8.9h)
of observing time, i.e. more than the other three band
together. Numbers are taken from the Subaru telescope's 
Suprime-Cam\footnote{Exposure Time Calculator at http://www.subaru.naoj.org/}.

\section{Using SN magnitude and color to estimate redshift}
Similar to using host galaxy magnitudes and colors, it is
possible to use these quantities of the SNe themselves to estimate the
redshift and SN type. For example, \cite{sn97ff} concluded
that the most distant known supernova, SN1997ff, was a Type Ia
supernova at a redshift $z=1.7 \pm 0.1$, based on the SN 
photometry. From the host galaxy, Riess et al derived a 
consistent photometric redshift, $z=1.65 \pm 0.15$.

One difficulty is the temporal evolution of the
SN light: one does not know at which phase in the lightcurve the SN
was discovered.  In this section we use the Monte-Carlo simulation
package SNOC \citep{snoc} to compute the SN discovery and selection
probabilities using the photometric and spectroscopic templates of
``normal'' Type Ia SNe described in \cite{nug01}. A potential
selection bias due to color-brightness correlations could be a source
of concern but is not considered in this work.

\begin{figure}
\figurenum{9}
\epsscale{0.8}
\plotone{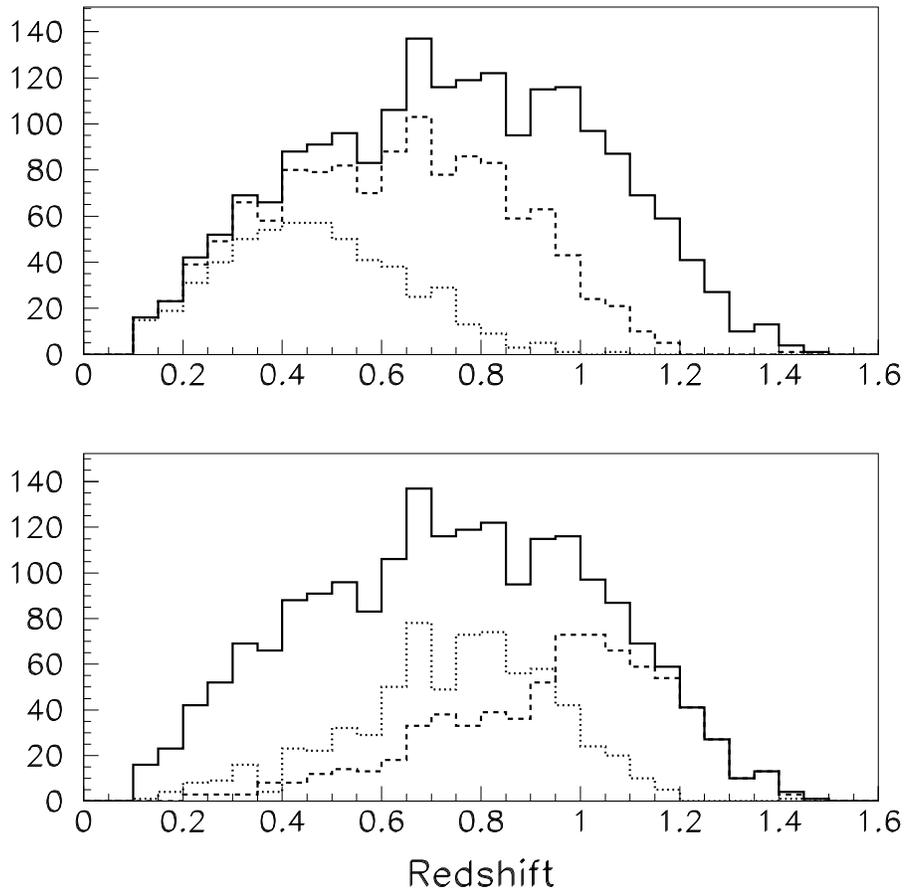}
\caption{The redshift distribution of discovered Type Ia SNe (in arbitrary units), assuming a constant SN rate per comoving volume. A discovery requires that the subtracted SN photometry between the detection and reference images exceeds the threshold brightness in the $I$-band. On the left
hand side, the redshift distributions are calculated for limiting
$I$-band magnitudes, $m_I$ = 23 (dotted), 24 (dashed) and 25 (solid),
using the SNOC Monte-Carlo package \citep{snoc}.  The bottom panel
shows also the redshift distribution that arises with
magnitude bounds on both sides: $23~<~I~<~24$ (dotted) and
$24~<~I~<~25$ (dashed).}
\label{fig9}
\end{figure}

The simulated redshift distribution of the discovered Type Ia SNe in
an $I$-band search is shown in the upper panel of Figure
\ref{fig9}. To produce a conservative estimate of the number of SNe with
redshift $z > 1$, we have assumed that the SN rate per comoving
volume is constant up to $z~\sim~1.5$.

In the simulations, SNe are selected when the subtraction
of SN photometry three weeks apart exceeds the threshold brightness
in I-band, e.g. $m_I~<~25$\footnote{The selection criteria is thus not identical
to the one described in section 2}. The absolute peak magnitude of the Type Ia
SNe is assumed to be $M_B=-19.33$, with an intrinsic spread of $\Delta
M_B^{rest} = 0.40 $ mag. This large spread gives a conservative 
representation of the distribution
of brightness for Ia supernovae prior to any lightcurve shape corrections. 
The intrinsic brightness spread at discovery is further 
enlarged  by the wide range of possible supernova lightcurve
phases.

For simplicity, we have assumed that all the
SNe have the same lightcurve shape, i.e. constant ``stretch'' (or
$\Delta m_{15}$). 

\begin{figure}
\figurenum{10}
\epsscale{0.8}
\plotone{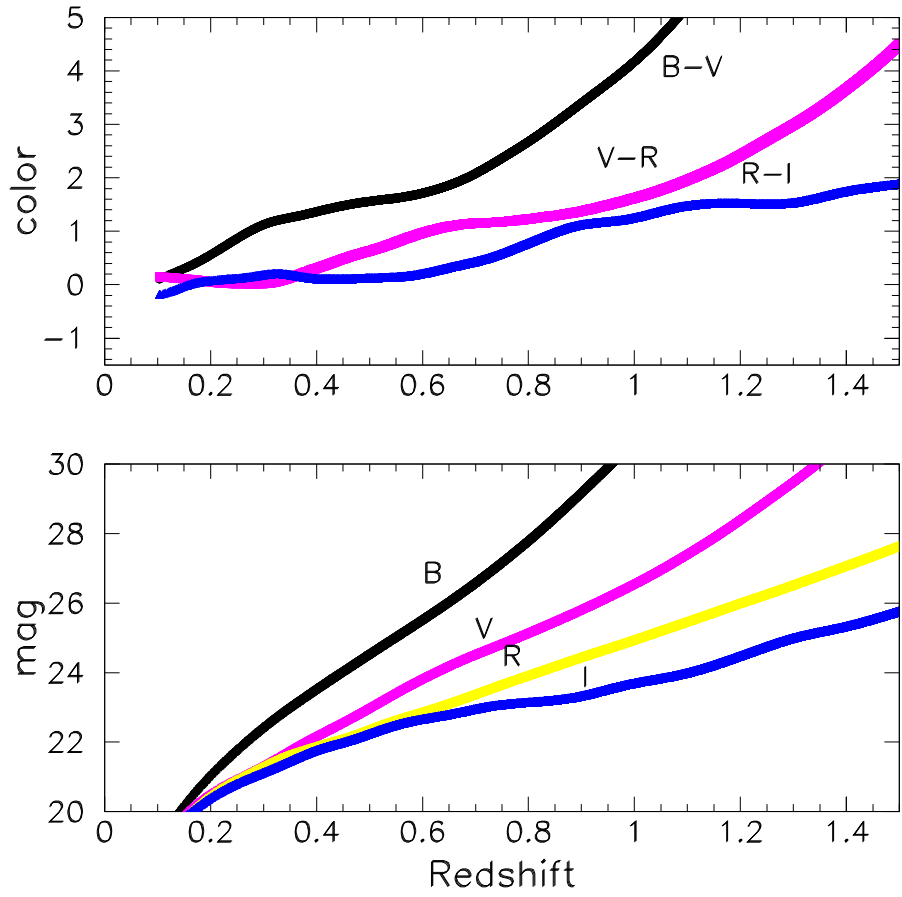}
\caption{Color and magnitude vs redshift at light curve maximum for ``normal'' Type Ia SNe.}
\label{fig10}
\end{figure}

As discussed earlier, Type Ia SNe are found in a very broad redshift
range if the selection of candidates is set by a threshold on the SN
$I$-band magnitude alone. One way to narrow down the distribution is
to impose a lower cut-off on the candidate's brightness. However, this
may not be sufficiently efficient as shown in the bottom panel of
Figure \ref{fig9}, where we plot the redshift distribution of SN in
two magnitude intervals, $23~<~m_I~<~24$ and $24~<~m_I~<~25$. Long
tails towards the lower redshifts makes this approach not optimal.  In
addition, this sort of criteria is likely to introduce a selection
bias which could severely worsen the precision of cosmological
parameters to be estimated from the data set and the measurements of
the SN rates. Next, we investigate the brightness and colors of Type
Ia SN vs redshift. Figure \ref{fig10} shows the optical color and
magnitude evolution at the date of maximum for the redshifted
rest-frame $B$-band light of a ``normal'' Type Ia SN.

\begin{figure}
\figurenum{11}
\epsscale{1.0}
\plottwo{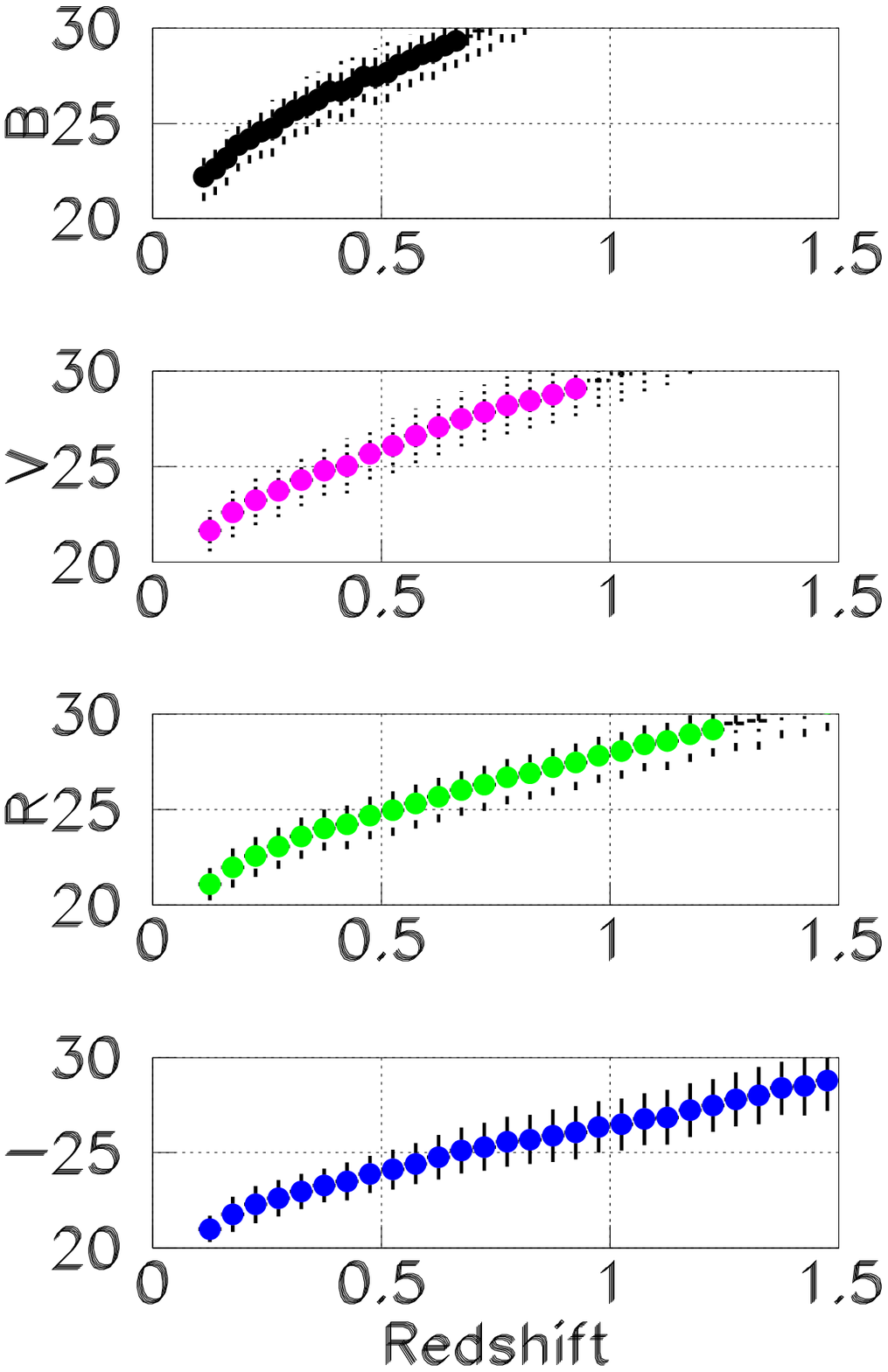}{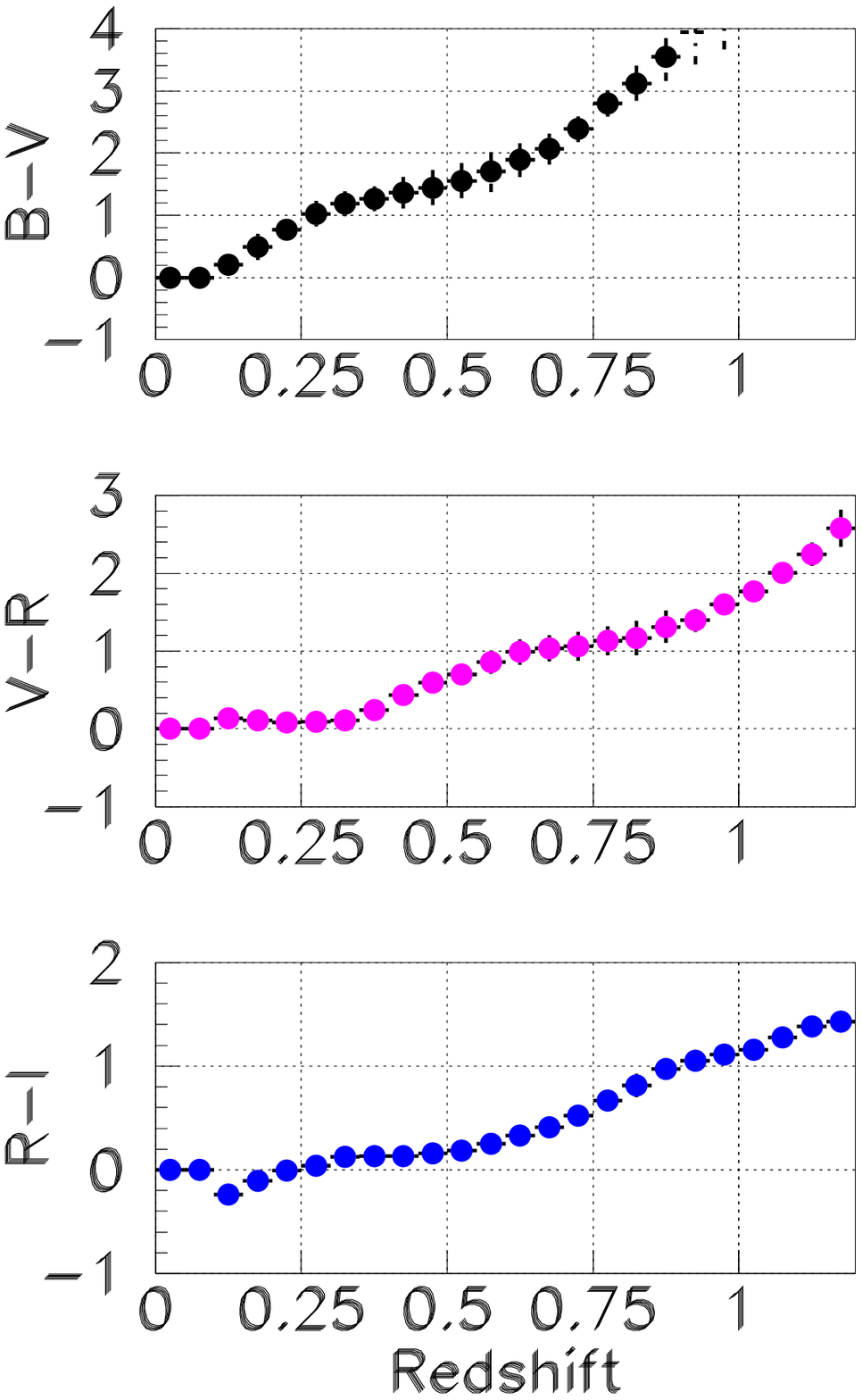}
\caption{Left: Average $BVRI$ magnitudes (top to bottom) vs redshift at the discovery epoch.
Right: average color vs redshift at discovery epoch for SN magnitudes $m_I<25$.} 
\label{fig11}
\end{figure}

We use SNOC to study the magnitudes and colors of the SNe at the
discovery epoch assuming a three weeks gap between reference and
discovery images. Figure \ref{fig11} shows the {\em average} magnitude
and its standard deviation in $BVRI$ and the $B-V$, $V-R$ and $R-I$
color at the discovery epoch. 
Note that the intrinsic spread in the SN peak magnitude, as well as the 
measurement errors,  are small compared to the magnitude range that emerges 
from all the possible reference and discovery epochs for the SNe. 

\subsection{Two and three band selection}
\begin{figure}
\figurenum{12}
\epsscale{1.0}
\plotone{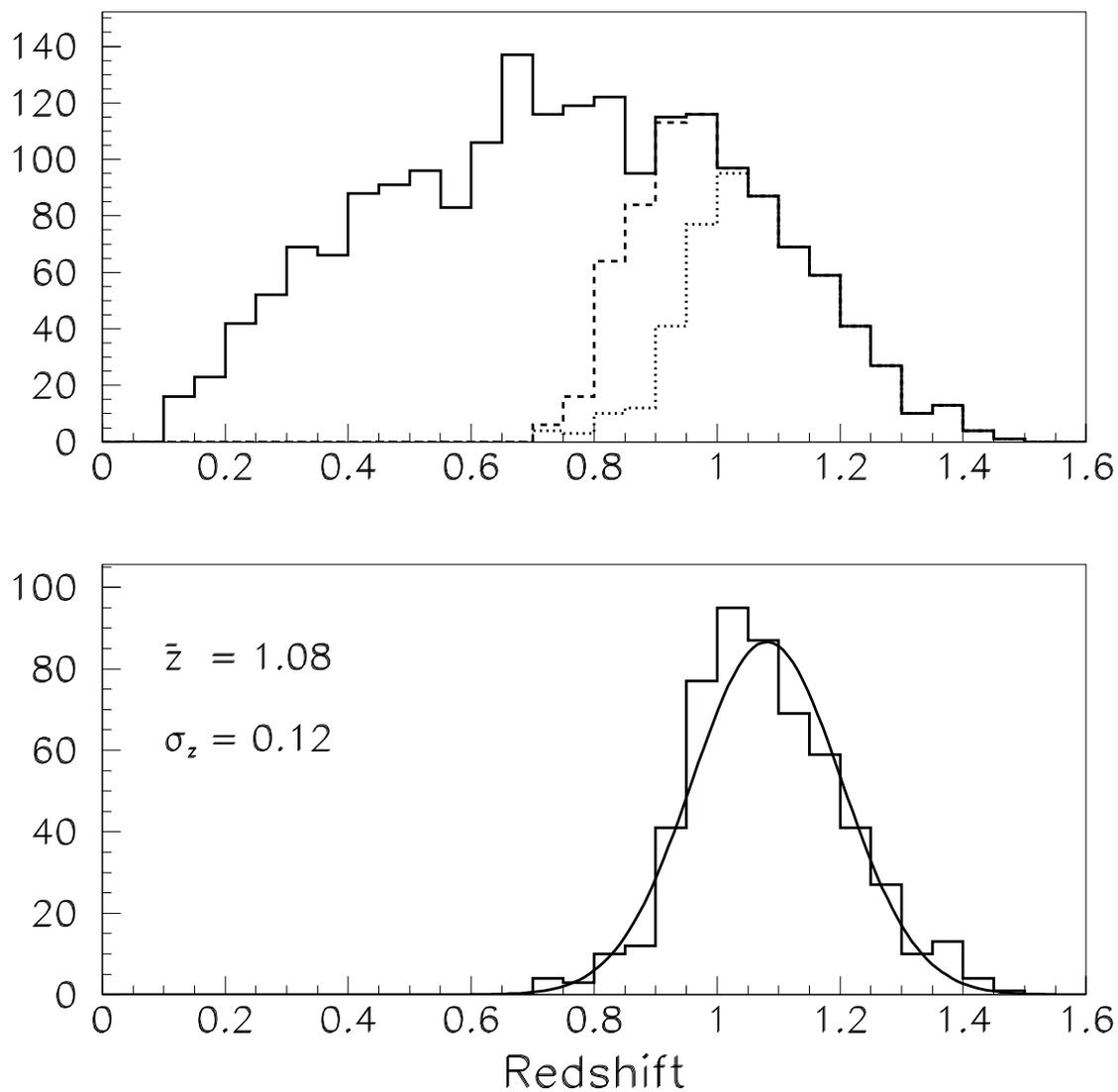}
\caption{Top panel: selected redshift range with cut on $R-I~>~0.8$ (dashed), and a combined criteria $R-I~>~0.8$ and $V-R~>~1.5$ (dotted). 
Bottom panel: Gaussian fit to the $z$-distribution imposing both selection criteria. The mean redshift of the selected SNe is $\bar{z}~\sim~1.1$, with a deviation $\sigma_z~\sim~0.13$ for an assumed color uncertainty (intrinsic + measure ment error of 0.2 magnitudes.}
\label{fig12}
\end{figure}

Next we show examples of possible selection criteria
using two and three filters. Figure \ref{fig12} shows the selected
range of redshifts demanding $R-I~>~0.8$ (dashed line) and {\em in
addition} demanding $V-I~>~1.5$ (dotted line). The solid line shows
the total distribution of Type Ia SNe with $m_I~\le~25$. We assume an
uncertainty in the color from both intrinsic color dispersion and
measurement Gaussian error of $\sigma_M$ = 0.1 magnitudes. In the bottom panel
we fit a Gaussian probability density function to the redshift distribution of 
SNe after imposing both criteria. This fit shows that it is possible to select
candidates around $z~\sim~1$ with an accuracy $\sigma_z \approx 0.12$, if
three bands are used. The $VRI$ selection rejects about 73 \% of the detected
SNe, but it is 82 (99)\% efficient in keeping candidates with $z\ge 0.9$
($z\ge 1.0$). Out of the selected sample, less than 6 (27) \% are below $z=0.9$ ($z=1.0$).

If the selection is done based solely on $R-I$, 41 \% of the detected
SNe are kept and the width of redshift
distribution is $\sigma_z \approx 0.15$. The  selection efficiency becomes nearly
100 \% for $z \ge 0.9$. The fraction of selected SNe below $z=0.9$
($z=1.0$) is then 21 (49) \%. Thus, the V-band information does help in 
rejecting the lower redshift SNe tail while introducing a moderate
decrease in overall efficiency. Given the faintness of $z > 1$ SNe
in $V$-band, $V~\gsim~27$ mag, a simple $R-I$ selection might be 
adequate in a realistic scenario.

\begin{figure}
\figurenum{13}
\epsscale{1.0}
\plotone{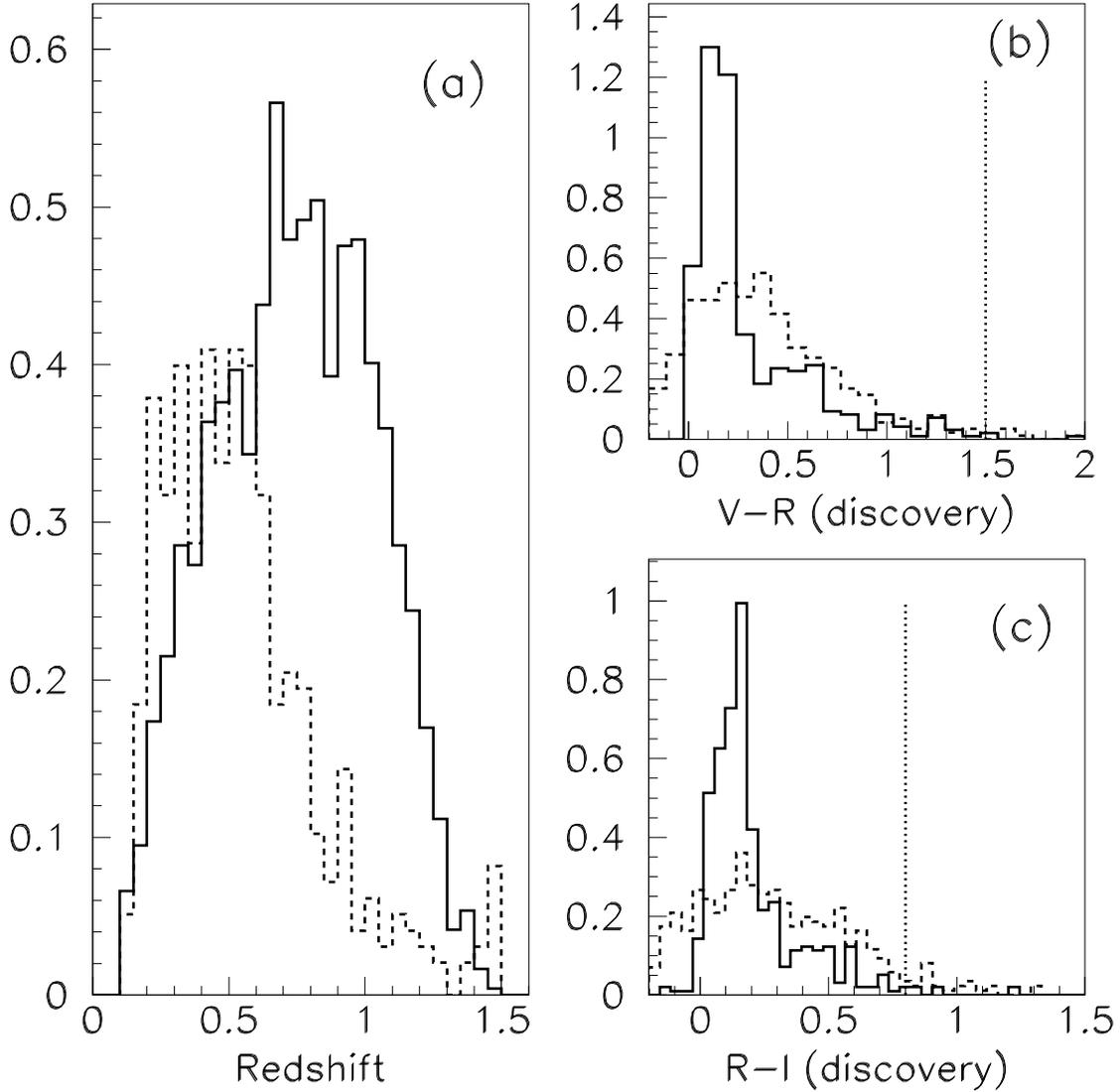}
\caption{(a) The redshift distribution of observable Type Ia SNe
(solid line) is compared to the non-Ia contamination (dashed line) for
a SN search with limiting magnitude $m_I < 25$ (units are
arbitrary). (b) The V-R color at the discovery epoch for the selected
non-Ia SNe. The dashed curves includes a 0.3 magnitude uncertainty from 
either measurement error or intrinsic dispersion. The dotted line shows the adopted color selection $V-R>
1.5$. (c) R-I at the discovery epoch for the non-Ia simulated SNe. The
dotted line shows the selection criteria $R-I> 0.8$}
\label{fig13}
\end{figure}

Finally, we address the issue of non-Ia contamination. Figure
\ref{fig13}a shows the simulated redshift distribution of discovered
Type Ia (solid line) and core collapse (dashed line) SNe for a three
week image separation search with limiting magnitude $m_I < 25$.  The
non-Ia SNe were simulated in the same way as Type Ia SNe. All SN types
were assumed to have constant rate per comoving volume in the
considered redshift range. The adopted relative intrinsic rate was
$N_{Ia}/N_{\rm core} = 1/2.6$ for $z~<~1.6$. Within the non-Ia class,
the various SN-types (Ib/c,IIL,IIn,IIP,87a-like) were given absolute
peak magnitudes, dispersions and relative fractions according to Table
1 in \cite{dah99}. For dust extinction by the SN host galaxy we assume
a face on value $A_{B0}$ = 0.32. Inclination effects are taken into
account by adding an extinction $A_{Bi}$ = 0.32$[(cosi)^{-1}-1]$,
where a random inclination, $i$, is assigned to each object.

The type of search we simulated produces a non-Ia candidate for every 
2 Type Ia supernovae found\footnote{This number is rather sensitive to
the limiting magnitude adopted. For example, for a limiting magnitude $m_I<23$ 
one expects only 1 non-Ia for every 5 Ia's assuming constant rate per comoving
volume for all SN types}.
Figures \ref{fig13}b and  \ref{fig13}c
show how the adopted color selections, $V-R > 1.5$ and $R-I > 0.8$
(and $m_I<25$) discriminate against the non-Ia supernovae, leaving a negligible
contamination rate, even when a 0.3 magnitude uncertainty in color from 
either measurement or intrinsic dispersion is added (dashed curves).

\section{Summary and conclusions}
We have shown that deep wide-field SN searches will detect a number of
SNe over a broad redshift range. E.g., an optical search to
$m_I~\sim~25$ will find Type Ia SNe to $z~\sim~1.5$. In this paper we discuss
the use of magnitudes and colors of the SNe, as well as the host
galaxies, in order for a prompt selection of high-$z$ SN candidates,
for subsequent spectroscopic follow-up. We first discuss the use of
additional photometric information on the host galaxies. The advantage
of using the host galaxies for redshift estimates is that a catalog of
high-$z$ candidate galaxies can be constructed in advance, enabling a
direct selection once a SN is discovered.

Using the available magnitude information in the search filter gives a
first indication of the probability that a SN is at high $z$. Adding
broad-band filters increases the probability for a correct
selection. Due to the cosmological shift of galaxy spectra, galaxies
appear redder when they are at higher redshift (in optical to
$z~\sim~1$). Using the color from two filters, together with the
magnitude information, one can therefore set additional constraints on
the redshift of the galaxy. Note, however, that different galaxy types
have different colors, as well as evolution in color, which makes a
single color difficult to use. The selection efficiency increases if the galaxy type is known, however, one must remember that there is no one-to-one relation between morphological type and colors.

If observations of the host galaxies are available in three filters, then
it is possible to derive selection criteria from the position of the
galaxy in a color-color diagram. As an example, we show how such
criteria can be constructed to select host galaxies with
$z~>~0.85$. Using $B$, $R$ and $I$ photometry, we construct a
selection criteria so that the probability is 80\% that a galaxy has
$z~>~0.85$, and where $\sim~64-86$ \% of the total number of host
galaxies above this redshift are actually selected. By changing the
selection criteria it is possible to increase the probability that the
host galaxy is at high-$z$, this will, however, lead to a decrease of
the fraction of the total number of high-$z$ host galaxies that are selected.

With four or more available filters it is
possible to estimate photometric redshift of the host galaxies. This results in an estimate of the redshift for each
galaxy separately, as compared to the three filter case, where we
instead calculate the probability that a host galaxy has a redshift
above, or below, some chosen limit.

We also discuss the necessary depth of a search in order to detect
host galaxies at $z~\sim~1$, and the depths required in other filters
in order to calculate colors. From the results listed in Table
\ref{Table1} \& \ref{Table3}, we see that with a limiting magnitude
$m_I~=~25$, about 80\% the host galaxies at $z~\sim~1.0$ can be
detected, while at $z~\sim~1.3$ the fraction is $\sim$ 60\%. The
required depths in the other optical filters to detect 90\% of the
galaxies with $m_I~=~25$ are $m_B~\sim~28.1$, $m_V~\sim~26.8$ and
$m_R~\sim~25.6$.

In a similar way, we show that colors of the Type Ia SNe can also be
used to constrain the redshift. A selection criteria based on $(V)RI$
magnitudes in a SN search with limiting magnitude $m_I~\sim~25$, leads
to a possible narrowing of the redshift range of the candidates to
$\sigma_z \le 0.15$ at $z~\sim~1$. Here 30-40\% of the total number of SNe are selected, of which 75-90\% have $z~>~0.9$. Multiband photometry of SNe 
is also very useful to distinguish between SNe of different types.
    
Selections based on the host galaxy magnitude and colors are 
preferable as they are less likely to introduce selection
biases. Also, obtaining colors of the SNe requires deep
multi-band  observations of each candidate taken within a
relatively short time interval, making this technique less
feasible.  

More observational data is
needed in order to understand and quantify the possible
color-brightness correlation of high-$z$ SNe and the resulting
selection bias.

Finally, the accuracy of the redshift determination, both 
form the galaxy and SN photometric information, depends on several parameters:
depths, number of available filters, accuracy in photometry etc, and
is a trade-off between required accuracy and the amount of observing
time available.


\acknowledgments
We wish to thank the anonymous referee for valuable comments and suggestions. We also thank Peter Nugent for providing us with spectral templates of Type Ia SNe. We are grateful to Saul Perlmutter and G\"{o}ran \"{O}stlin for useful discussions. The work by TD is supported by the Swedish Research Council and
the Swedish Board for Space Sciences. AG is a Royal Swedish Academy Research Fellow supported by a grant from the Knut and Alice Wallenberg Foundation.




\clearpage


\clearpage 

\clearpage

\clearpage

\clearpage

\clearpage

\clearpage

\clearpage

\clearpage

\clearpage

\clearpage

\clearpage

\clearpage





\clearpage
\begin{table}
\caption{Rest frame $B$ magnitudes as a function of redshift for five different galaxy types, assuming an {\it observed magnitude} $m_I~=~25$. The three rightmost columns list the fraction of the SN host galaxies that can be detected at the different redshifts for three limiting magnitudes, $m_I~=~24$, $m_I~=~25$ and $m_I~=~26$.}
\begin{tabular}{ccccccccc}
\tableline
\tableline
$z$ & E & Sa & Sbc & Scd & Im & $f(m_I=24)$ & $f(m_I=25)$ & $f(m_I=26)$ \\
\tableline
0.8 &     -17.5  &  -17.5 & -17.5  & -17.5 & -17.5 & 0.80 & 0.91 & 0.96\\
0.9 &     -18.1  &  -18.0 & -17.9  & -17.9 & -17.8 & 0.73 & 0.87 & 0.94\\ 
1.0 &     -18.8  &  -18.6 & -18.4  & -18.3 & -18.1 & 0.63 & 0.82 & 0.92\\
1.1 &     -19.3  &  -19.2 & -18.8  & -18.7 & -18.5 & 0.51 & 0.75 & 0.88\\ 
1.2 &     -19.8  &  -19.7 & -19.3  & -19.1 & -18.9 & 0.40 & 0.67 & 0.84\\ 
1.3 &     -20.1  &  -20.1 & -19.7  & -19.5 & -19.2 & 0.30 & 0.58 & 0.79\\ 
1.4 &     -20.5  &  -20.4 & -20.1  & -19.8 & -19.4 & 0.22 & 0.50 & 0.74\\ 
\tableline
\end{tabular}
\label{Table1}
\end{table}

\clearpage
\begin{table}
\caption{Apparent magnitudes in different filters for galaxies at different redshifts, when the {\em observed magnitude} is $m_I$ = 25.}
\begin{tabular}{ccccccccccccc}
\tableline
\tableline
\multicolumn{1}{c}{} & \multicolumn{3}{c}{Elliptical}& \multicolumn{3}{c}{Sbc galaxy}& \multicolumn{3}{c}{Irr galaxy}\\ 
\tableline
z    & $m_B$ & $m_V$ & $m_R$ & $m_B$ & $m_V$ & $m_R$ & $m_B$ & $m_V$ & $m_R$\\ 
0.8  & 29.9  & 27.7  & 26.6  & 27.9  & 27.1  & 26.2  & 26.6  & 26.3  & 25.8 \\
0.9  & 29.8  & 27.8  & 26.5  & 27.8  & 27.1  & 26.3  & 26.5  & 26.2  & 25.8 \\
1.0  & 29.6  & 27.9  & 26.4  & 27.7  & 27.1  & 26.3  & 26.4  & 26.2  & 25.8 \\
1.1  & 29.4  & 28.1  & 26.4  & 27.5  & 26.9  & 26.3  & 26.3  & 26.0  & 25.7 \\
1.2  & 29.2  & 28.2  & 26.5  & 27.3  & 26.8  & 26.2  & 26.1  & 25.9  & 25.6 \\
1.3  & 29.2  & 28.4  & 26.8  & 27.2  & 26.6  & 26.1  & 26.0  & 25.8  & 25.5 \\
1.4  & 29.1  & 28.3  & 26.9  & 27.0  & 26.5  & 26.0  & 26.0  & 25.7  & 25.4 \\
\tableline
\end{tabular}
\label{Table2}
\end{table}

\clearpage
\begin{center}
\begin{table}
\caption{Limiting magnitudes in $B$, $V$ and $R$ that are required in order to detect 70, 80, 90 and 95\% of the galaxies with $m_I~\lsim~25$.}
\begin{tabular}{ccccccccccccc}
\tableline  
\tableline  
   &\multicolumn{3}{c}{$f_{70}$} & \multicolumn{3}{c}{$f_{80}$} & \multicolumn{3}{c}{$f_{90}$} & \multicolumn{3}{c}{$f_{95}$} \\
\tableline  
\hline
 z       & $m_B$ & $m_V$ & $m_R$ & $m_B$ & $m_V$ & $m_R$ & $m_B$ & $m_V$ & $m_R$ & $m_B$ & $m_V$ & $m_R$\\ 
0.8 & 26.1 & 25.0 & 24.3 & 26.7 & 25.5 & 24.7 & 27.5 & 26.0 & 25.2 & 28.1 & 26.4 & 25.6\\
0.9 & 26.3 & 25.4 & 24.6 & 27.0 & 25.8 & 25.0 & 27.9 & 26.3 & 25.5 & 28.5 & 26.8 & 25.8\\
1.0 & 26.4 & 25.7 & 24.9 & 27.2 & 26.1 & 25.2 & 28.1 & 26.6 & 25.6 & 28.6 & 27.1 & 25.9\\
1.1 & 26.3 & 25.8 & 25.1 & 27.3 & 26.3 & 25.4 & 28.2 & 27.0 & 25.7 & 28.7 & 27.4 & 26.0\\
1.2 & 26.2 & 25.8 & 25.2 & 27.2 & 26.4 & 25.5 & 28.2 & 27.2 & 25.8 & 28.7 & 27.7 & 26.1\\
1.3 & 26.0 & 25.7 & 25.3 & 27.0 & 26.3 & 25.6 & 28.2 & 27.4 & 26.0 & 28.8 & 27.9 & 26.3\\
1.4 & 25.9 & 25.6 & 25.3 & 26.7 & 26.1 & 25.5 & 28.1 & 27.3 & 26.2 & 28.7 & 27.9 & 26.6\\ \tableline
\end{tabular}
\label{Table3}
\end{table}
\end{center}



\begin{thebibliography}{}

\bibitem[Bolzonella, Miralles \& Pell\'{o}(2000)]{bol00}
Bolzonella M., Miralles J.--M., \& Pell\'{o} R. 2000, \aap, 363, 476

\bibitem[Brunner, Connolly \& Szalay(1999)]{bru99}
Brunner R.J., Connolly A.J., \& Szalay A.S. 1999, \apj, 516, 563

\bibitem[Cappellaro et al.(1993)]{cap93}
Cappellaro E., et al. 1993, \aap, 268, 472  

\bibitem[Cohen et al.(2000)]{coh00}
Cohen, J. G., et al. 2000, ApJ, 538, 29

\bibitem[Coleman, Wu \& Weedman(1980)]{col80}
Coleman G.D., Wu C.-C., \& Weedman D.W. 1980, \apjs, 43, 393

\bibitem[Connolly et al.(1995)]{con95}
Connolly A.J., Csabai I., Szalay A.S., Koo D.C., Kron R.G., \& Munn J.A. 1995, \aj, 110, 2655

\bibitem[Connolly et al.(1997)]{con97}
Connolly A.J., Szalay A.S., Dickinson M., SubbaRao M.U., \& Brunner R.J. 1997, \apjl, 486, 11

\bibitem[Dahl\'{e}n \& Fransson(1999)]{dah99} Dahl\'{e}n T., \& Fransson C. 1999, \aap, 350, 349

\bibitem[Dressler et al. (1999)]{dr99} Dressler A., Smail I., Poggianti B.M., Butcher H., Couch W.J., Ellis R.S., Oemler A. 1999, \apjs, 122, 51

\bibitem[Fern\'{a}ndes-Soto et al.(1999)]{fer99}
Fern\'{a}ndes-Soto A., Lanzetta K.M., \& Yahil A. 1999, \aj, 513, 34

\bibitem[Fontana et al.(2000)]{fon00}
Fontana A., et al. 2000, \aj, 120, 2206

\bibitem[Goliath et al.(2001)]{gol01}
Goliath M., Amanullah R., Astier P., Goobar A., \& Pain R. 2001,  
\aap, 380, 6

\bibitem[Goobar \& Perlmutter(1995)]{goo95}
Goobar A., \& Perlmutter S. 1995, \apj, 450, 14

\bibitem[Goobar et al.(2001)]{snoc}
Goobar A., et al., 2001, in preparation

\bibitem[Gwyn(1995)]{gwy95}
Gwyn S. 1995, MS Thesis, University of Victoria

\bibitem[He, Zou \& Zhang(2001)]{he01}
He P., Zou Z.-L., \& Zhang Y.-Z. 2001, preprint, astro-ph/0101095

\bibitem[Kinney et al.(1996)]{kin96}
Kinney A.L., Calzetti D., Bohlin R.C., McQuade K., Storchi--Bergmann T., \& Schmitt H.R. 1996, \apj, 467, 38

\bibitem[Lilly et al.(1995)]{lil95}
Lilly S.J., Tresse L., Hammer F., Crampton D., \& Le F\`{e}vre O. 1995, \apj, 455, 108

\bibitem[Madau, Della Valle \& Panagia(1998)]{mad98a}
Madau P., Della Valle M., \& Panagia N. 1998, \mnras, 297, L17

\bibitem[Madau, Pozzetti \& Dickinson(1998)]{mad98b}
Madau P., Pozzetti L., \& Dickinson M. 1998, \apj, 498, 106

\bibitem[Metcalfe et al.(1991)]{met91}
Metcalfe N., Shanks T., Fong R., \& Jones L.R. 1991, \mnras, 249, 498

\bibitem[Mobasher et al.(1996)]{mob96}
Mobasher B., Rowan--Robinson M., Georgakakis A., \& Eaton N. 1996, \mnras, 282, L7

\bibitem[Nugent et al.(2001)]{nug01}
Nugent P., et al. 2001, {\em submitted to \pasp}

\bibitem[Perlmutter et al.(1999)]{per99}
Perlmutter S., et al. 1999, \apj, 517, 565

\bibitem[Poggianti et al. (1999)]{po99} Poggianti B.M., Smail I., Dressler A., Couch W.J., Berger A.J., Butcher H., Ellis R.S., Oemler A. 1999, \apj, 518, 576

\bibitem[Puschell, Owen \& Laing(1982)]{pus82}
Puschell J.J., Owen F.N., \& Laing R.A. 1982, \apj, 275, 57

\bibitem[Riess et al.(1998)]{rie98}
Riess A.G., et al. 1998, \aj, 116, 1009

\bibitem[Riess et al.(2001)]{sn97ff}
Riess A.G., et al. 2001, \apj, 560, 49

\bibitem[Sawicki, Lin \& Yee(1997)]{saw97}
Sawicki M.J., Lin H., \& Yee H.K.C. 1997, \aj, 113, 1

\bibitem[Schechter(1976)]{sch76}
Schechter P. 1976, \apj, 203, 297

\bibitem[Wang, H\"{o}flich \& Wheeler(1997)]{wan97}
Wang L., H\"{o}flich P., \& Wheeler J.C. 1997, \apj, 483, L29

\bibitem[Wang, Bahcall \& Turner(1998)]{wan98}
Wang Y., Bahcall N., \& Turner E.L. 1998, \aj, 116, 2081

\end{thebibliography}
\end{document}